\documentstyle[epsf]{aa}
\begin{document}

\newcommand{\gsim}{\hbox{\rlap{$^>$}$_\sim$}}
%
   \title{Images, Light Curves and Spectra of GRB Afterglow}

   \author{Jonathan Granot  \inst{1}, Tsvi Piran \inst{1,2} \and Re'em Sari  \inst{3}
          }

   \institute{Racah Institute of Physics, The Hebrew University, 
              Jerusalem 91904, Israel. \\ 
       \and  Department of Physics, Columbia University, 
              New York, NY 10027, USA \\
\and  Theoretical Astrophysics, California Institute of Technology, 
              Pasadena, CA 91125, USA.}
   \date{Received December 15, 1998; accepted} 

   \maketitle

   \begin{abstract}
     We calculate the light curve and spectra near the peak and the self
     absorption break, for an adiabatic blast wave described by the
     Blandford-McKee solution, considering the emission from the whole
     region behind the shock front. The expected light curve and
     spectra are flat near the peak. This rules out the interpretation
     of the sharp peak observed in the optical afterglow of GRB970508
     as the expected peak of the light curve. The observed image of an
     afterglow is calculated for a broad range of frequencies. We show
     that for frequencies below the self absorption frequency the
     image is rather homogeneous, as opposed to the bright ring at the
     outer edge and dim center, which appear at higher frequencies.
     We fit the observed spectra of GRB970508 to the detailed theory
     and obtain estimates of the physical parameters of this burst.

   \end{abstract}

%

\section{The Physical Model}
We consider emission from the whole volume behind an adiabatic highly
relativistic spherical blast wave expanding into a cold and uniform
medium. The hydrodynamics is described by the Blandford-McKee (1976
denoted BM hereafter) self similar solution.  
For typical parameters, the evolution becomes adiabatic fairly
early, about an hour after the initial burst (\cite{SPN}, Granot,
Piran \& Sari 1998a and 1998b, hereafter GPSa and GPSb, respectively).
The BM solution is valid from this time, and as long as 
$\gamma \gsim 2$ (Kobayashi et. al. 1998), typically a few months after the burst.

We assume that $\nu_a \ll \nu_m$, where $\nu_m$ is the peak frequency
and $\nu_a$ is the self absorption frequency, which is reasonable for
the first few months. The dominant radiation emission mechanism is
assumed to be synchrotron radiation, while Compton scattering and
electron cooling are ignored. We denote quantities measured in the
local rest frame of the matter with a prime, while quantities without
a prime are measured in the observer frame.

We assume that the energy of the electrons is everywhere a constant
fraction of the internal energy: $e'_{el}=\epsilon_e e'$, and consider
a power law electron distribution: $N(\gamma_e)\propto\gamma_e^{-p}$
for $\gamma_e \ge \gamma_{min}$. The magnetic field is also assumed to
hold a constant fraction of the internal energy: $e'_{B}=\epsilon_B
e'$, where $e_B=B^2/8\pi$ is the energy density of the magnetic field.
Alternative magnetic field models were considered in GPSa and GPSb,
and we obtained that our results are not sensitive to the assumptions
on the magnetic field.

\section{The Observed Image}

   \begin{figure}

     \epsfxsize=8.8cm \epsfbox{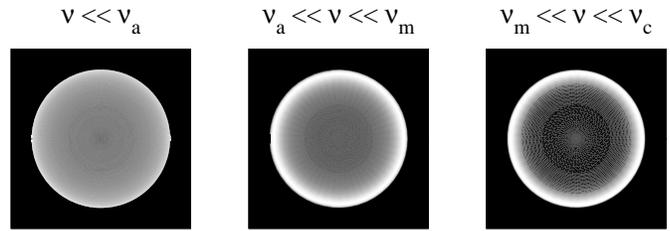}
      \caption[]{The observed image of a GRB afterglow at a given
        observed time, for different frequencies. 
              }
         \label{images}
   \end{figure}
   
   The observed images, at various
   frequencies, are shown in Figure \ref{images} (GPSa, GPSb). For $\nu
   \gg \nu_{m}$ a thin bright ring appears on the outer edge of the image,
   while the center is much dimmer (only a few percent of the maximal
   surface brightness). For $\nu_a \ll \nu \ll \nu_m$ the surface
   brightness at the center is $34\%$ of the maximal surface
   brightness, and $58\%$ of the average surface brightness.
   For $\nu \ll \nu_a$ the surface brightness at the
   center of the image is $77\%$ of its average value, resulting in an
   almost uniform disk.

\section{Light Curve, Spectra \& The Burst Parameters}
\begin{figure}

     \epsfxsize=8.8cm \epsfbox{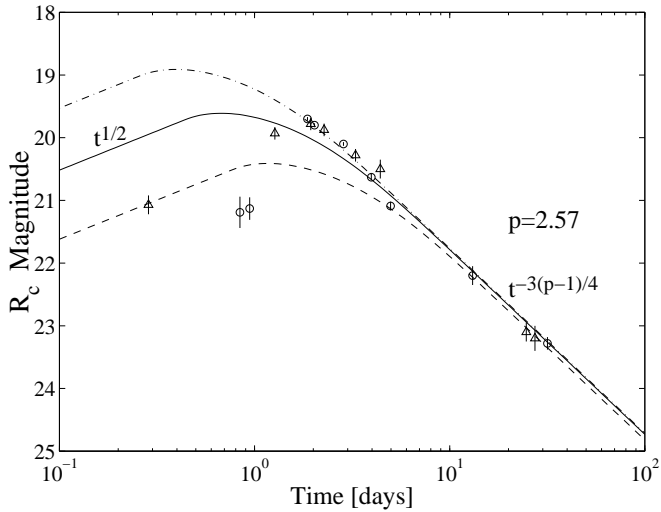}
      \caption[]{Optical 
        observations of GRB970508, made by Sokolov et al. (1997, circles) and
        Metzger et al. (1997, triangles). The three curves are three
        possible theoretical light curves. }
         \label{light_curve}
   \end{figure}

The light curve and spectra of an afterglow are flat near the peak.
The exact shape, and the value of the peak frequency and peak flux
depend on the values of the physical parameters of the burst (see
GPSa). In Figure \ref{light_curve} we see the peak in the optical
light curve of GRB970508 (Sokolov et al. 1997, Metzger et al. 1997),
with three theoretical light curves. These light curves are for
$p=2.57$, which corresponds to the power law decay that follows the
peak, and differ by the values of the remaining parameters.

It is quit evident from Figure \ref{light_curve} that the shape of the 
optical peak displayed by GRB970508 is not accounted for by the
theoretical light curve arising from the model we used, and a
different explanation should be considered.

\begin{figure}
  
  \epsfxsize=8.8cm \epsfbox{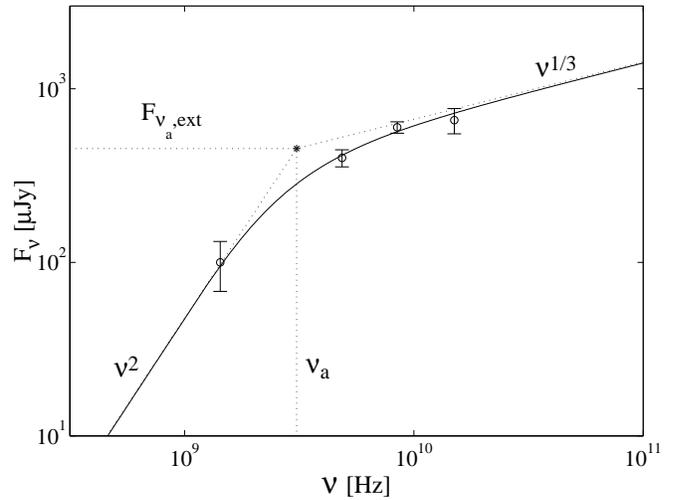}
      \caption[]{A fit of our calculated spectra to radio afterglow observations 
        of GRB970508, made about one week after the burst.
              }
         \label{spectra}
   \end{figure}

The observed flux density near the self absorption frequency (i.e., 
$\nu \ll \nu_m$) can be approximated, with an accuracy better than
3\%, by the following simple expression:
\begin{equation}
\label{disk}
F_{\nu}=F_{\nu_a,ext}\psi^2\left( 1-\exp[-\psi^{-5/3}]\right)
\quad,\quad \psi \equiv \nu/\nu_a,
\end{equation}
Where $F_{\nu_a,ext}$ and $\nu_a$ are defined in Figure \ref{spectra}
and depend on the values of the physical parameters (GPSb).

In Figure \ref{spectra} we show a fit between our theoretical spectra
near $\nu_a$ and radio afterglow observations of GRB970508 (Shepherd
et al. 1998), which were made about one week after the
burst. Extracting the values of  $F_{\nu_a,ext}$ and $\nu_a$ from this
fit and comparing them to the theoretical values results in two
constraints on the parameters of the burst (GPSb). 

Using the values of $\nu_a, \nu_m, F_{\nu_m}$ and the cooling break
frequency $\nu_c$ extracted from the broad band spectra of GRB970508,
and comparing them to their theoretical values, Wijers \& Galama
(1998) calculated the parameters of this burst, using a simple broken
power law theoretical spectra (\cite{SPN}). Making a similar
calculation, using the more detailed description of the self absorption
break and of the spectral peak, we obtained the following values 
for the physical parameters of GRB970508:
\begin{eqnarray}
\label{phys_con}
\quad \quad \quad \quad E=&5.3\times 10^{51} \ {\rm ergs} \quad \quad
\ \ \epsilon_e&=0.57 \\ \nonumber
\quad \quad \quad \quad n_0=&5.3 \ {\rm cm^{-3}} \quad  \quad \quad
\quad \ \ 
\epsilon_B&=0.0082 \ \ ,
\end{eqnarray}
where $E$ is the total energy of the shell and $n_0$ is the ambient
number density.

\section{Discussion}
We have calculated the light curve and spectra due to synchrotron
emission from an extreme relativistic adiabatic blast wave, which is
described by the Blandford-McKee (1976) self similar solution. We
obtained a flat peak for the light curve, which rules out the
interpretation of the sharp optical peak of GRB970508 as the expected
peak of the light curve. The observed image of an afterglow has been
calculated over a wide range of frequencies. The image at $\nu<\nu_a$
is quite homogeneous, while at higher frequencies there is a bright
ring at the outer edge and the center is dim.

We have calculated the physical parameters of GRB970508 based on
detailed calculations of the spectra near $\nu_a$ and $\nu_m$.
The values we have obtained are different by up to two orders of 
magnitude from the values obtained
using a simple broken power law theoretical spectra (\cite{W&G}). This
stresses the sensitivity of this method to the exact details of the
theoretical model.

\begin{acknowledgements}
This research was supported by the US-Israel BSF grant 95-328 and
by a grant from the Israeli Space Agency. Re'em Sari thanks the
Sherman Fairchild Foundation for support.
\end{acknowledgements}

\end{document}